
\font\mybb=msbm10 at 12pt
\def\bb#1{\hbox{\mybb#1}}
\def\bZ {\bb{Z}}
\def\bR {\bb{R}}
\def\bE {\bb{E}}
\def\bT {\bb{T}}
\def\bM {\bb{M}}


\tolerance=10000
\input phyzzx

\input epsf
\ifx\epsfbox\UnDeFiNeD\message{(NO epsf.tex, FIGURES WILL BE
IGNORED)}
\def\figin#1{\vskip2in}
\else\message{(FIGURES WILL BE INCLUDED)}\def\figin#1{#1}\fi
\def\ifig#1#2#3{\xdef#1{fig.~\the\figno}
\goodbreak\midinsert\figin{\centerline{#3}}%
\smallskip\centerline{\vbox{\baselineskip12pt
\advance\hsize by -1truein\noindent\footnotefont{
}}}
\bigskip\endinsert\global\advance\figno by1}

\def\footnotefont{\tenpoint}

\newwrite\ffile\global\newcount\figno \global\figno=1
\def\fig{fig.~\the\figno\nfig}
\def\nfig#1{\xdef#1{fig.~\the\figno}%
\writedef{#1\leftbracket fig.\noexpand~\the\figno}%
\ifnum\figno=1\immediate\openout\ffile=figs.tmp\fi\chardef\wfile=
\ffile%
\immediate\write\ffile{\noexpand\medskip\noexpand\item{Fig.\
\the\figno. }
\reflabeL{#1\hskip.55in}\pctsign}\global\advance\figno by1\findarg}

 \def\unit{\hbox to 3.3pt{\hskip1.3pt \vrule height 7pt width 
.4pt \hskip.7pt
\vrule height 7.85pt width .4pt \kern-2.4pt
\hrulefill \kern-3pt
\raise 4pt\hbox{\char'40}}}

\REF\wit{E. Witten, {\it Some comments on string dynamics}, 
hep-th/9507121, to
appear in the proceedings of {\it Strings '95}, USC, March 1995.}
\REF\ver{E. Verlinde, Nucl. Phys. {\bf B445} (1995) 211.}
\REF\tow{P.K. Townsend, Phys. Lett. {\bf 139B} (1984) 283; 
M. Awada, G. Sierra and
P.K. Townsend, Class. Quantum Grav. {\bf 2} (1985) L85.}
\REF\dl{M.J. Duff and J.X. Lu, Nucl. Phys. {\bf B411} (1994) 301.}
\REF\memdyon{J.M. Izquierdo, N.D. Lambert, G. Papadopoulos and 
P.K. Townsend,
Nucl. Phys. {\bf B460}, (1996) 560.}
\REF\jhsa{J.H. Schwarz, Phys. Lett. {\bf 360B} (1995) 13.}
\REF\jhsb{J.H. Schwarz, Phys. Lett. {\bf 367B} (1996) 97.}
\REF\aha{O. Aharony, {\it String theory dualities from M-theory},
hep-th/9604103.}
\REF\gp{G. Papadopoulos, {\it A brief guide to p-branes}, 
hep-th/9604068.}
\REF\giv{A. Giveon and M. Porrati, {\it Aspects of 
Space-Time Dualities},
hep-th/9605118.}
\REF\gib{G.W. Gibbons, Nucl. Phys. {\bf B207} (1982) 337.}
\REF\ghs{D. Garfinkle, G. Horowitz and A. Strominger, Phys. 
Rev. {\bf D43} (1991) 3140.}
\REF\tsey{A. Tseytlin, {\it Harmonic superpositions of M-branes}, hep-th/9604035.}
\REF\pt{G. Papadopoulos and P.K. Townsend, {\it Intersecting 
M-branes}, hep-th/9603087.}
\REF\witb{E. Witten, Nucl. Phys. {\bf B460} (1996) 335.}


\Pubnum{ \vbox{ \hbox{DAMTP-R/96/26} } }
\pubtype{}
\date{May 1996}

\titlepage

\title {\bf Dyonic p-branes from self-dual (p+1)-branes}

\author{M.B. Green, N.D. Lambert, G. Papadopoulos and P.K. Townsend}
\address{DAMTP, Silver St., 
\break
Cambridge CB3 9EW, U.K.}

\abstract{The `electromagnetic' $Sl(2;\bZ)$ duality group 
in spacetime dimension $D=4k$ can be given a Kaluza-Klein 
interpretation in $D=4k+2$ as the modular group of a compactifying 
torus. We show how dyonic $2(k-1)$-branes
in $D=4k$ can be interpreted as self-dual $(2k-1)$-branes 
in $D=4k+2$ wound around the homology cycles of the torus. 
In particular, dyons of the D=4 N=4 heterotic string
theory are interpreted as winding modes of a D=6 self-dual 
string, while D=8 dyonic membranes are interpreted as wound 
3-branes of D=10 IIB superstring theory. We also
discuss the T-dual IIA interpretations of D=8 dyonic membranes.}

\endpage


\chapter{Introduction}

It is now widely appreciated that extended objects, or p-branes, 
play an important part in our understanding of superstring 
dualities in various spacetime dimensions D. This paper is 
dedicated to a study of the role of dyonic p-branes in
spacetime dimensions $D=4k$, for integer $k$. The prototype is 
that of D=4 dyons of the $\bT^6$-compactified heterotic string, 
and there is now strong evidence that
they fill out multiplets of the S-duality group $Sl(2;\bZ)$. 
It was observed in [\wit] that this group can be interpreted 
as the modular group of the Kaluza-Klein (KK) torus arising from compactification on $\bT^2$ of a new 
self-dual D=6 superstring theory; a similar observation was made
in [\ver] in the context of an abelian gauge theory. In the 
context of the  effective D=6 field theory the self-dual string 
can be identified as a string soliton
of the chiral N=4 supergravity obtained by $K_3$-compactification 
of D=10 IIB supergravity [\tow]; it is just the self-dual string 
soliton found in [\dl] as a solution of  N=2 D=6 supergravity. 
One result of this paper is to show that the winding modes of this 
D=6 self-dual string around the homology cycles of the
KK torus can be identified as D=4 dyons filling out S-duality 
multiplets. 

In [\wit] the D=6 self-dual string theory was argued to result 
from the $K_3$-compactification of the D=10 type IIB superstring 
theory, the self-dual string itself being identified  as a IIB 
3-brane wrapped around a two-cycle of $K_3$. 
Thus, S-duality in string theory was shown to be a consequence 
of the non-perturbative equivalence of the $\bT^6$-compactified 
heterotic string theory to the $K_3\times \bT^2$ compactified 
IIB superstring theory. In this `10 to 6 and then to 4' approach 
to heterotic S-duality, D=6 is merely a convenient
intermediate dimension arrived at by consideration of the 
$K_3$ compactification {\it prior} to a subsequent $\bT^2$ 
compactification. Clearly, one could perform the
$\bT^2$ compactification first to arrive at an intermediate 
type II D=8 superstring theory. This `10 to 8 and then to 4' 
approach was explored in [\memdyon], except that the D=10 starting 
point was there taken to be the IIA superstring theory. Here
we concentrate on the IIB case, although we shall have more to 
say about IIA at the conclusion of the paper. 

The D=8 type II superstring theory has an $Sl(3;\bZ)\times 
Sl(2;\bZ)$ U-duality group, with the $Sl(2;\bZ)$ subgroup acting 
by `electromagnetic' duality on the 4-form field strength 
(descending from the 4-form field strength of D=11
supergravity) whose sources are dyonic membranes. As we shall see, 
this $Sl(2;\bZ)$ `electromagnetic' duality group (which is also 
a subgroup of the $Sl(2;\bZ)\times Sl(2;\bZ)$ T-duality group) 
can be identified as the modular group of the torus
arising from compactification to D=8 of the D=10 type IIB 
superstring, and the dyonic membranes as windings of the IIB 
3-brane around the homology cycles of the torus. As explained 
in [\memdyon], upon further $K_3$-compactification the D=8 dyonic 
membranes can be wrapped around the homology 2-cycles of $K_3$ to 
yield D=4 dyons which can be identified as those filling out
S-duality multiplets in the equivalent heterotic string.

The two approaches to the interpretation of D=4 heterotic S-duality 
in terms of the D=10 IIB superstring theory are illustrated by the 
following diagram of commuting compactifications:
\vskip 0.5cm

\ifig\ffour{   }
{\epsfbox{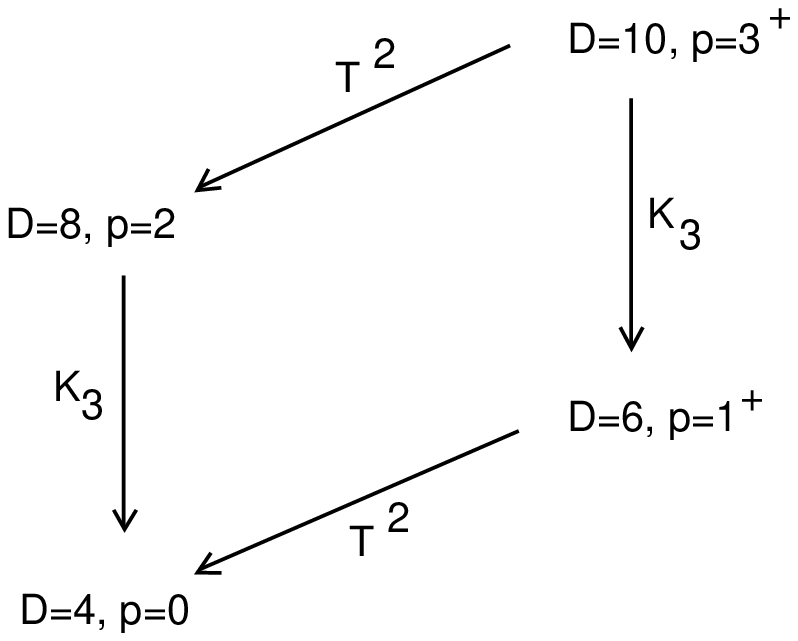}}

Whether one adopts the `10 to 6 and then to 4' or the `10 to 8 and 
then to 4' point of view, there is an $Sl(2;\bZ)$ subgroup of the 
U-duality group in $D=4k$ dimensions, for either $k=1$ or $k=2$, 
that can be interpreted as the modular group of the Kaluza-Klein 
torus for a $\bT^2$ compactification from $D= 4k+2$ dimensions. 
The main purpose of this paper is to show that a dyonic
p-brane of the effective $(4k)$-dimensional field theory with 
(magnetic, electric) charge vector $(m,n)$ can be interpreted as a $(4k+2)$-dimensional self-dual (p+1)-brane wrapped $m$ times around 
one homology cycle of the torus and $n$ times around the other. 
Since the basic idea is applicable in spacetime
dimensions $D=4k$ for arbitrary integer $k$ we shall first 
develop it in this more general context. 

Specifically, we consider an $Sl(2;\bR)$-invariant
$4k$-dimensional field theory for a $(2k)$-form field strength 
$F$, and a complex scalar field $\tau$ taking values in 
$Sl(2;\bZ)\backslash SL(2,\bR)/U(1)$. We shall show that this theory
is the dimensional reduction on $\bT^2$ (and consistent 
truncation) of a self-dual $(2k+1)$-form theory in a 
$(4k+2)$-dimensional spacetime $M_{4k+2}= M_{4k}\times
\bT^2$, with the modular `parameter' of $\bT^2$ identified 
with the complex scalar field of the $D=4k$ theory and the 
modular group of $\bT^2$ identified with the
$SL(2,\bZ)$ duality group of the $D=4k$ theory.
We then apply this result for $k=1,2$ to show, firstly, that 
dyons of D=4 N=4 supergravity can be lifted to D=6 as winding 
modes of a D=6 self-dual string and, secondly, that the dyonic 
membrane solutions of D=8 N=2 supergravity can be lifted to
D=10 as winding modes of the self-dual IIB 3-brane. These 
results are reminiscent of the  way in which the $Sl(2;\bZ)$ 
multiplet of IIB strings [\jhsa] can be understood
as winding modes of the D=11 membrane via a compactification 
to D=9 on $\bT^2$ [\jhsb], although the details are somewhat 
different. 

It was shown in [\memdyon] that the D=8 dyonic membrane solutions 
can be lifted to D=11 to give new solutions of D=11 supergravity 
that interpolate between the membrane and 5-brane\foot{Thus, D=11 membrane/5-brane duality can be viewed as a manifestation of 
`electromagnetic' duality in D=8; an elaboration of this point 
can be found in [\aha].}. These solutions, which also preserve 
half the supersymmetry, might be viewed as the effective field theory realisation of membrane/5-brane bound states. Alternatively, the 
D=8 dyonic membrane solutions can be lifted to D=10 to yield solutions 
of IIA supergravity that might represent membrane/4-brane bound states. Remarkably, these membrane/4-brane solutions preserve half the 
supersymmetry and are {\it self-dual}. They can be found by 
reduction from D=11, which is the method used here to find them. 
We also use this method to find two other D=10 `bound state' solutions, originally suggested in [\gp], which turn out to
be duals of each other.

During the completion of this work a paper appeared [\giv] 
having some overlap with section 2 of this paper.


\chapter{$\bT^2$-reduction of self-dual gauge theories}

Let $F=dA$ be a $(2k)$-form field strength and $\tau$ a complex
scalar field in a $(4k)$-dimensional spacetime, $M_{4k}$. 
Consider the action
$$
I_{4k}= \int d^{4k}x\; \sqrt{-g}\bigg[ R -{1\over2} 
{d\tau d\bar\tau\over ({\rm Im}\,
\tau)^2}\bigg]
\ + \  \alpha \!\int \!\! F\wedge G .
\eqn\eone
$$
where 
$$
G={\rm Im}\tau\, {}^*F-{\rm Re}\tau\, F
\eqn\efour
$$
and $\alpha$ is a real constant; for agreement with the 
conventions of [\memdyon] we must choose
$$
\alpha = 2[(2k)!]^{-2}\ .
\eqn\aeone
$$
The asterisk in \efour\ indicates the Hodge dual with respect 
to the metric on $M_{4k}$.

The kinetic term for $\tau$ is invariant under the 
$Sl(2;\bR)$ transformation
$$
\tau \rightarrow {a\tau+b\over c\tau+d}\ .
\eqn\eonea
$$
The $A$-field equations are also $SL(2,\bR)$-invariant 
provided that $F$ transforms by `electromagnetic' duality, i.e.
$$
(F,G)\rightarrow (F,G) A^{-1}\ ,
\eqn\ethree
$$
where $A$ is the $SL(2, \bR)$ matrix
$$
A=\pmatrix{a & b\cr c &d}\ .
\eqn\efive
$$

We now show how the action \eone\ can be obtained by 
compactification on $\bT^2$ of a $(4k+2)$-dimensional action. 
We do so by generalizing the method of [\ver] to apply
to curved space backgrounds with Minkowski signature. We start 
from the $D=4k+2$ action
$$
I_{(4k+2)}= \int\! d^{(4k+2)}x\; \sqrt{-g} R \ +\  
\alpha\! \int\!\!  \big[
dC\wedge H+{1\over 2} H\wedge{}^*H\big]\ ,
\eqn\esix
$$
where $C$ is a $(2k)$-form and $H$ is a $(2k+1)$-form on a
$(4k+2)$-dimensional spacetime $M_{(4k+2)}$. The matter 
field equations are
$$
\eqalign{
H-{}^*dC&=0
\cr
dH&=0\ ,}
\eqn\eseven
$$
which imply the standard free field equation for $C$. The 
matter part of the action  \esix\ is simply the first order 
formulation of the Maxwell-type gauge theory for
the $2k$-form $C$. This formulation will prove convenient 
for the later incorporation of a self-duality condition. 

We now consider the compactification of \esix\ on a torus 
$\bT^2$, i.e. $M_{4k+2}=M_{4k}\times \bT^2$. Let $(x,y)$ be 
coordinates for $\bT^2$ with
standard identifications, i.e
$$
x \sim x+ 1\qquad y\sim y+1\ ;
\eqn\iden
$$
the shape of the torus is then determined by its modular 
parameter, $\tau$, which will be a function on $M_{4k}$ in the 
KK ansatz. This ansatz (which includes a consistent truncation) is 
$$
\eqalign{
ds^2 (M_{4k+2})&=ds^2 (M_{4k})+ 
{1\over {\rm Im}\tau}\big(|\tau|^2 dy^2+2{\rm Re} 
\tau\; dx dy+ dx^2\big)\cr
H&= G dy+F dx 
\cr
C&= B dy+ A dx \ ,}
\eqn\enine
$$
where $F,G$ are $(2k)$-forms on $M_{4k}$, while $A,B$ are 
$(2k-1)$-forms and $\tau$ is a complex scalar field. A feature of 
this KK ansatz is that the volume of the torus is unity. More 
generally one could take the volume to be arbitrary, which
would introduce an additional scalar field on $M_{4k}$ in the 
KK ansatz. Effectively we have set this field to the constant 
corresponding to unit volume. This is a consistent truncation 
provided that we keep only that part of the gauge field action
which is conformally invariant in $4k$ dimensions, as we have done, 
because the volume field of the torus does not
couple to the conformally invariant part of the action. 

The $4k$-dimensional action resulting from the above KK ansatz is
$$
\eqalign{
I_{(4k)} =\int d^{4k}x\; &\sqrt{-g}\big[ R -{1\over2} 
{d\tau d\bar\tau\over ({\rm Im}
\tau)^2}\big] +\alpha \int \bigg\{ \big(dA\wedge G-
dB\wedge F\big)\cr
& +{1\over 2\, {\rm Im}\tau} \big(G\wedge {}^*G+2 
{\rm Re}\,\tau F\wedge {}^*G +
|\tau|^2 F\wedge {}^*F\big)\bigg\}\ . }
\eqn\eten
$$
Following [\ver] we now take into account a self-duality 
condition on the
original field-strength $H$ by imposing the constraint
$$
i_V(H-dC)=0\ ,
\eqn\eeleven
$$
where $V={\partial/\partial x}$. From the KK ansatz we 
see that this gives
$$
F=dA\ .
\eqn\etwelve
$$

The $2k$-form $G$ in \eten\ is auxiliary and may be eliminated 
by its field equation. Taking into account that $F=dA$ this is 
just $G={\rm Im} \tau\, {}^*F-{\rm Re} \tau
F$, as in \efour. Using this in \eten, and dropping the
$dB\wedge F$ term which is now a total derivative, we get 
precisely the action \eone.  As a check, we observe that if we 
substitute for $G$ in terms of $F$ in the KK ansatz \enine, we 
discover that the $(2k+1)$-form field strength $H$ is self-dual.


\chapter{Dyonic p-branes from self-dual (p+1)-branes}

We now turn to the applications of the reduction procedure 
just described for $k=1$ and $k=2$. For $k=1$ the action \esix\ is 
a consistent truncation of the N=4 chiral
D=6 supergravity theory (which is itself a truncation of the 
anomaly-free effective N=4 chiral supergravity obtained by 
compactification of IIB supergravity on $K_3$).
Using the KK ansatz \enine\ and setting 
$$
\tau=2 \rho+i e^{-2\sigma}\ , 
\eqn\dthree
$$ 
we find the D=4 action
$$
\eqalign{
I = \int \Bigg\{\sqrt{-g}\big[ &R - 2\partial_\mu 
\sigma\partial^\mu\sigma -
2e^{4\sigma}\partial_\mu \rho\partial^\mu\rho  -
e^{-2\sigma}F_{\alpha\beta}F^{\alpha\beta}\big] \cr
&-\varepsilon^{\mu\nu\alpha\beta}\rho 
F_{\mu\nu}F_{\alpha\beta} \Bigg\}\ .}
\eqn\sdone
$$
where $F_{\mu\nu}$ are the components of the 2-form field 
strength $F_2$.

The field equations of this action have the following 
extreme charged black hole
solutions [\gib,\ghs]:
$$
\eqalign{
ds_{(4)}^2=&-H^{-1} dt^2 + H\, ds^2(\bE^3)
\cr
F_{2}=& {1\over\sqrt{2}}\bigg[ \cos\xi (\star dH)+ 
\sin\xi\, dH^{-1}\wedge dt\bigg]
\cr
\tau=& { \sin(2\xi) (1-H^2)+2 i H \over 2 
(\sin^2\xi+H^2 \cos^2\xi)}\ ,}
\eqn\sdseven
$$
where $\xi$ is an angle, $H$ is a harmonic function on 
$\bE^3$ and $\star$ is the Hodge star of $\bE^3$. We have chosen 
the asymptotic value $<\tau>$ of the complex scalar field $\tau$ 
to be $i$. The angle $\xi$ is the parameter of a $U(1)$
rotation of the purely magnetic solution. The existence of 
this one-parameter family of `classical dyons' follows from 
the $Sl(2;\bR)$ invariance of the action and the existence of 
a $U(1)$ subgroup of $Sl(2;\bR)$ that leaves invariant the 
asymptotic value $<\tau>$, the $U(1)$ subgroup depending on 
$<\tau>$. In the quantum theory this $U(1)$ group is broken to 
$\bZ_2$ by the Dirac quantization condition. In particular,
when $<\tau>=i$ the only surviving values of $\xi$ are those 
for which $\cos\xi=0,1$. It might appear from this that the 
classical dyon solution \sdseven\ is irrelevant in the quantum 
theory. This would indeed be true if the scalar field
target space ${\cal M}$ were the homogeneous space 
$Sl(2;\bR)/U(1)$, as is the case {\it locally}, but in the string 
theory context S-duality implies that
$$
{\cal M} = Sl(2;\bZ)\backslash Sl(2;\bR)/U(1)\ .
\eqn\target
$$
This means that the asymptotic value $<\tau>=i$ should not 
be distinguished from any of the other values obtained from it 
by an $Sl(2;\bZ)$ transformation, but it follows from \ethree\ 
that such a transformation takes a purely magnetic
solution, with (magnetic, electric) charge vector $(1,0)$ 
to a dyon with (magnetic, electric) charge vector $(d,-b)$ where 
$a,b,c,d$ are {\it integer} entries of the
matrix $A$ of \efive\ satisfying $ad-bc=1$ (so that now 
$A\in Sl(2;\bZ)$). 

To obtain the D=6 interpretation of these dyons we shall need 
the explicit form of the $Sl(2;\bZ)$ transformed solution. 
Starting from the $\xi=0$ case of \sdseven\
an $Sl(2;\bZ)$ transformation leads to 
$$
\eqalign{
ds_{(4)}^2=&-H^{-1} dt^2 + H\, ds^2(\bE^3)
\cr
F_{2}=& {1\over \sqrt{2}}e^{<\sigma>}\big[\cos\psi 
\star dH + \sin\psi\, dH^{-1}\wedge dt\big] \cr
\tau =& 2<\rho> + e^{-2<\sigma>}\bigg[ {\sin(2\psi)(1-H^2) + 
2iH\over 2(\sin^2\psi +
H^2\cos^2\psi)}\bigg]\ ,}
\eqn\sdsevena
$$
where $H$ is a harmonic function on $\bE^3$. The asymptotic 
value of $\tau$ is now
$$
<\tau> = {bd+ac + i\over c^2+d^2}\ ,
\eqn\asymtau
$$
and 
$$
\tan\psi =  c/d\ ,
\eqn\asymtan
$$
where $a,b,c,d$ are the integer entries of the $Sl(2;\bZ)$ 
matrix $A$. 

Referring to the KK ansatz \enine, one sees that the solution 
lifts to the D=6 self-dual string solution 
$$
\eqalign{
ds_{(6)}^2=&H^{-1} (-dt^2 + dv^2) + H\, [ds^2(\bE^3)+du^2]
\cr
F_{3}=& {1\over\sqrt{2}}\bigg[ (\star dH)\wedge du+  
dH^{-1}\wedge \epsilon(\bM^2) \wedge dv\bigg] \ ,}
\eqn\sdeight
$$
where $\epsilon(\bM^2)$ is the volume form on the two-dimensional 
Minkowski spacetime $\bM^2$ with coordinates $(t,v)$, provided 
that the coordinates $(u,v)$ are related to $(x,y)$ of the KK 
ansatz by
$$
(y,x) = (v,u)A^{-1}\ .
\eqn\dnine
$$
The configuration \sdeight\ is precisely the self-dual string 
solution of D=6 N=4 supergravity with a choice of harmonic 
function consistent with the KK ansatz.
Clearly, $v$ must be identified with the string's spatial 
coordinate. We can choose to identify it with unit period, 
so that $\oint\! dv =1$ where the integral is over one period. 
The winding numbers of the string around the $x$
and $y$ directions of the torus $\bT^2$ are then
$$
\eqalign{
\big(\oint\! dy, \oint\! dx \big) &= 
\big(\oint\! dv{\partial y\over\partial v}, 
\oint\! dv{\partial x\over\partial v} \big)\cr
&= (d,-b)\ , }
\eqn\winding
$$
which is also the (magnetic, electric) charge vector of the 
D=4 dyon.

The solution \sdseven\ is a special case of a more general 
extreme dyonic black hole depending on two harmonic functions 
[\gib], but this more general solution preserves only $1/4$ of 
the N=4 supersymmetry and lifts to a self-dual 
string solution preserving only $1/4$ of the supersymmetry of 
N=4 chiral D=6 supergravity. Such a D=6 string cannot be 
interpreted as a IIB 3-brane wrapped around a homology cycle of 
$K_3$ because the latter configuration preserves
$1/2$ the supersymmetry. However, a D=6 self-dual string preserving 
$1/4$ of the N=4 supersymmetry has a D=10 interpretation as two 
IIB 3-branes intersecting on a string [\tsey], with the four `relative transverse dimensions' (in the terminology of [\pt]) compactified 
on two 2-cycles of $K_3$. We shall not have anything
further to say here about dyons preserving $1/4$ of the 
supersymmetry except to point out that their existence is directly 
related to the existence of an N=2 supergravity, in which 
context they preserve $1/2$ the
supersymmetry. This is to be contrasted with the situation 
pertaining to D=8 dyonic membranes, to be discussed below. 
Since the minimal D=8 supergravity
theory admitting supersymmetric dyonic membranes 
is actually maximally supersymmetric, the general 
supersymmetric dyonic membrane can depend on only
one harmonic function. We now turn to the D=10 IIB 
interpretation of these solutions.

For $k=2$ the action \esix\ is, if the equations of 
motion are augmented by the self-duality constraint, a consistent 
truncation of the action of D=10 IIB supergravity. Applying the 
reduction procedure and setting $\tau=2 \rho+i
e^{-2\sigma}$, as before, we obtain the following D=8 action:
$$
\eqalign{
I = \int \Bigg\{\sqrt{-g}\big[ &R - 2\partial_\mu 
\sigma\partial^\mu\sigma -
2e^{4\sigma}\partial_\mu \rho\partial^\mu\rho  
-{1\over
12}e^{-2\sigma}F_{\alpha\beta\gamma\delta}
F^{\alpha\beta\gamma\delta}\big] \cr
&-{1\over 144}\varepsilon^{\mu\nu\rho\sigma\alpha\beta\gamma\delta}
\rho F_{\mu\nu\rho\sigma}F_{\alpha\beta\gamma\delta} \Bigg\}\ .}
\eqn\done
$$
This is just the action used in [\memdyon], where the field 
equations were shown to
admit the following `classical' dyonic membrane solutions:
$$
\eqalign{
ds_{(8)}^2=&H^{-{1\over2}} ds^2(\bM^3) + H^{1\over 2} ds^2(\bE^5)
\cr
F_{4}=& {1\over2} \cos\xi (\star dH)+{1\over2} \sin\xi dH^{-1}\wedge
\epsilon(\bM^3)
\cr
\tau=& { \sin(2\xi) (1-H)+2 i H^{1/2}\over 2 
(\sin^2\xi+H \cos^2\xi)}\ ,}
\eqn\dseven
$$
where $ds^2(\bM^3)$ is the 3-dimensional Minkowski metric, 
$\epsilon(\bM^3)$ is the volume form of $\bM^3$, $\xi$ is an 
angle, $H$ is a harmonic function on $\bE^5$ and
star is the Hodge star in $\bE^5$. 

To find the D=8 dyonic membranes relevant to the quantum 
theory we proceed as for the D=4 dyons. We first set $\cos\xi=1$ 
thereby selecting the purely magnetic solution [\dl], to which we 
assign (magnetic, electric) charge vector $(1,0)$,
and we then perform an $Sl(2;\bZ)$ transformation to arrive at a 
solution with (magnetic, electric) charge vector $(d,-b)$ for 
co-prime integers $b,d$. The $Sl(2;\bZ)$-transformed solution has 
the same (Einstein) metric as in \dseven, but the other fields are 
now 
$$
\eqalign{
F_{4}=& {1\over2}e^{<\sigma>}\big[ \cos\psi(\star dH) +
\sin\psi\, dH^{-1}\wedge\epsilon(\bM^3)\big]\cr
\tau=& 2<\rho> + e^{-2<\sigma>}\, .\,
{\sin(2\psi) (1-H)+2 i H^{1/2}\over 2 (\sin^2\psi +H \cos^2\psi)}\ ,}
\eqn\dsevena
$$
where the vacuum values of $\rho$ and $\sigma$ are given by 
\asymtau\ and the angle $\psi$ is again given by \asymtan. 

Using the ansatz \enine\ for $k=2$ we deduce that these 
dyonic membrane solutions lift to the following solutions in D=10:
$$
\eqalign{
ds_{(10)}^2=&H^{-{1\over2}} (ds^2(\bM^3)+dv^2) + 
H^{1\over 2} (ds^2(\bE^5)+du^2)
\cr
F_{5}=& {1\over2} (\star dH)\wedge du+{1\over2}  
dH^{-1}\wedge \epsilon(\bM^3)
\wedge dv\ ,}
\eqn\seight
$$
where the coordinates $(u,v)$ are again related to 
$(x,y)$ as in \dnine. The configuration \seight\ is precisely 
the self-dual 3-brane solution of IIB
supergravity except that $H$ is a harmonic function on 
$\bR^5$ rather than $\bR^6$ because of consistency with the KK 
ansatz. The winding numbers of this 3-brane around the KK torus, 
parameterized by $(x,y)$, is given by the same computation 
as before, with the result that a dyonic membrane with
(magnetic, electric) charge vector $(m,n)$ is a 3-brane wound 
$m$ times around one fundamental homology cycle of the KK 
torus and $n$ times around the other one.  


\chapter{Dyonic membranes as membrane-fourbrane bound states}

In section 3, we showed that the D=8 dyonic membrane solutions 
could be lifted to D=10 as solutions of IIB supergravity. We 
now turn to their IIA interpretation. This question was already 
partly addressed in [\memdyon] since it was shown there that 
the dyonic membrane solutions could be lifted to D=11 as new
solutions of D=11 supergravity interpolating between the 
membrane and the 5-brane solution. Such solutions can be viewed 
as special cases of orthogonally intersecting p-branes in which 
one brane lies entirely within the other. Thus the D=11 solution
under discussion could be interpreted as a membrane within a 
5-brane, i.e. $(2|2,5)$ in the notation of [\gp] where the last 
set of numbers indicate the values of $p$ for the intersecting 
$p$-branes and the first number gives the value of
$p$ for the common intersection. By lifting to D=10 rather 
than D=11 we instead find new solutions of the D=10 IIA 
supergravity that interpolate between the
membrane and the 4-brane solutions, i.e. $(2|2,4)$. Given
that the D=11 result is already known, it is actually simpler 
to reduce from D=11 rather than lift from D=8, and we shall 
adopt that procedure here. Thus, our starting
point will be the following D=11 $(2|2,5)$ solution [\memdyon]:
$$
\eqalign{
ds^2_{(11)}&= H^{-{2\over3}} [\sin^2\xi+H\cos^2\xi]^{1\over3} 
ds^2(\bM^3)+
H^{1\over3} [\sin^2\xi+H\cos^2\xi]^{1\over3} ds(\bE^3)
\cr
&+H^{1\over3}
[\sin^2\xi+H\cos^2\xi]^{-{2\over3}} ds^2(\bE^5)
\cr
F^{(11)}_{4}&={1\over2} \cos\xi \star dH+ 
{1\over2}\sin\xi dH^{-1}\wedge
\epsilon(\bM^3) \cr
& \qquad + {3 \sin 2\xi\over 2[\sin^2\xi+H\cos^2\xi]^2} 
dH \wedge
\epsilon(\bE^3)\ ,}
\eqn\iione
$$
where $H$ is a harmonic function on $\bE^5$, and  
$\epsilon(\bM^3)$ and $\epsilon(\bE^3)$ are the volume 
forms on $\bM^3$ and $\bE^3$, respectively.

We use the following KK ansatz for reducing, and truncating, D=11 
supergravity to D=10, with the D=10 metric in the string frame: 
$$
\eqalign{
ds^2_{(11)}&=e^{-{2\over3}\phi} ds^2_{(10)} +e^{{4\over3}\phi} du^2
\cr
F^{(11)}_4&=F_4+F_3\wedge du\ .}
\eqn\iitwo
$$
 
Using this ansatz there are still three distinct ways of 
reducing the D=11 $(2|2,5)$
solution to D=10, depending on the choice of compactifying 
direction. The corresponding D=10 solutions have the 
interpretation as a membrane within a 
4-brane $(2|2,4)$, a membrane within a 5-brane, $(2|2,5)$ 
and a string within a 4-brane $(1|1,4)$. As explained above, 
the case of immediate interest is the first,
and using the ansatz \iitwo\ we find the following new 
$(2|2,4)$ solution of D=10 IIA supergravity:
$$
\eqalign{
ds^2_{(10)}&=H^{-{1\over2}} ds^2(\bM^3)+{H^{1\over2}\over
\sin^2\xi+H\cos^2\xi} ds^2(\bE^2)+H^{1\over2} ds^2(\bE^5)
\cr
F_4&={1\over2} \cos\xi \star dH+{1\over2} \sin\xi\, 
dH^{-1}\wedge \epsilon(\bM^3)
\cr
F_3&= {3 \over 2}{\sin(2\xi)\over [\sin^2\xi+H\cos^2\xi]^2}
\epsilon(\bE^2)\wedge dH
\cr
e^{4\phi}&={H\over [\sin^2\xi+H\cos^2\xi]^2} \ .}
\eqn\iithree
$$
This configuration is presumably the field-theoretic 
realization of a IIA superstring  D-2-brane within a D-4-brane. 
Remarkably, the IIA $(2|2,5)$ solution is
self-dual in the sense that the procedure used in [\pt] to 
construct a dual solution yields the same solution.

We conclude by considering the other two possible reductions 
of the D=11 $(2|2,5)$ solution. Using the same KK ansatz but 
reducing in the other two ways, as explained
above, we find, firstly, the following  D=10 IIA $(1|1,4)$ solution:
$$
\eqalign{
ds^2_{(10)}&=H^{-1} [\sin^2\xi+H\cos^2\xi]^{1\over2}
ds^2(\bM^2)+[\sin^2\xi+H\cos^2\xi]^{-{1\over2}}
ds^2(\bE^3)
\cr
&+[\sin^2\xi+H\cos^2\xi]^{1\over2} ds^2(\bE^5)
\cr
F_4&={1\over2} \cos\xi \star dH-{3\over 2}{\sin(2\xi)\over
[\sin^2\xi+H\cos^2\xi]^2}\epsilon(\bE^3)\wedge dH
\cr
F_3&={1\over2} \sin\xi dH^{-1}\wedge \epsilon(\bM^2)
\cr
e^{4\phi}&={[\sin^2\xi +H\cos^2\xi]\over H^2}\ .}
\eqn\iithree
$$
This is presumably the field theoretic realization of a bound 
state of fundamental IIA string with a D-4-brane, perhaps 
analogous to the bound states of IIB fundamental
strings with D-1-branes discussed in [\witb].

Secondly, we find the following D=10 IIA $(2|2,5)$ solution:
$$
\eqalign{
ds^2_{(10)}&=H^{-{1\over2}} (\sin^2\xi+H\cos^2\xi)^{1\over2} 
ds^2(\bM^3)+
H^{{1\over2}} (\sin^2\xi+H\cos^2\xi)^{-{1\over2}} ds^2(\bE^3)
\cr
&+ H^{{1\over2}}
(\sin^2\xi+H\cos^2\xi)^{{1\over2}} ds^2(\bE^4)
\cr
F_4&=-{3\over2} {\sin(2\xi)\over [\sin^2\xi+H\cos^2\xi]^2}
\epsilon(\bE^3)\wedge dH+{1\over2}
\sin\xi dH^{-1}\wedge
\epsilon(\bM^3)
\cr
F_3&={1\over2} \cos\xi \star dH\ 
\cr
e^{4\phi}&=H [\sin^2\xi+H\cos^2\xi] \ .}
\eqn\iiithree
$$
This is just the magnetic dual of \iithree. 
\vskip 0.5cm

\noindent{\bf Acknowledgments:}  G.P. is supported by a 
University Research Fellowship from the Royal Society.

\refout

\bye